\documentclass[12pt]{article}
\usepackage{amssymb}
\usepackage{amsmath}
\usepackage[ps,dvips,matrix,arrow,frame,import,curve,color]{xy}
\makeatletter
\@addtoreset{equation}{section}
\makeatother

\def\be{\begin{equation}}
\def\ee{\end{equation}}

\def\ba{\begin{array}}
\def\ea{\end{array}}

\newcommand{\bea}{\begin{eqnarray}}
\newcommand{\eea}{\end{eqnarray}}
\def\N{$\cal N$}

\def\E {$E_{7(7)}$}

\begin{document}
\hfill{}

\begin{flushright}
\end{flushright}

\vskip 1cm

\vspace{24pt}

\begin{center}
{ \LARGE {\bf    The Ultraviolet Finiteness \\

\vskip 0.6cm

of   \N=8 Supergravity }}

\vspace{24pt}

{\large  {\bf     Renata Kallosh }}

    \vspace{15pt}

{ Department of Physics,
 Stanford University, Stanford, CA 94305}

\vspace{10pt}

\vspace{24pt}

\end{center}

\begin{abstract}

We study  counterterms (CT's), candidates for UV divergences in the four-dimensional  \N=8 supergravity.  They have been constructed long ago in the Lorentz covariant  on shell superspace and  recently in the chiral light-cone (LC) superspace. We prove  that all of these CT's  are ruled out  since they are not available in the real LC superspace. This implies the perturbative UV finiteness of $d=4$  \N=8 supergravity under the assumption that  supersymmetry and continuous \E \,symmetry are anomaly-free. The proof, based on the chiral nature of CT's in  the LC superspace, is a  generalization of the  perturbative F-term non-renormalization theorem for \N=8 supergravity.

\end{abstract}
\newpage
\newpage
\section{Introduction}
The story of  \N=8 supergravity \cite{Cremmer:1979up, Brink:1979nt} has been  developing rapidly during the last few years, as shown in review papers \cite{Dixon:2010gz}. The recently established properties are: it is UV finite up to the 4-loop level \cite{Dixon:2010gz} and it is likely to be all-loop anomaly free \cite{Bossard:2010dq}. The absence of chiral $SU(8)$ and continuous \E \,symmetry anomalies is rather non-trivial. It is crucial for the argument of this paper since it is based on the equivalence theorem between the Lorentz covariant and LC superspace path integrals of the theory in the form established in \cite{Kallosh:2009db}.

An infinite set of \E \, invariant  UV candidate CT's was constructed  in \cite{Howe:1980th, Kallosh:1980fi} using the geometric on shell Lorentz-covariant superspace \cite{Brink:1979nt} with 32 Grassmann coordinates.  Despite some differences concerning CT's and the onset of UV divergences in \cite{Kallosh:2009db,Kallosh:2008mq} and   \cite{Bossard:2009sy},   there seems to be a consensus that the infinite set of CT's  is still to be taken seriously.  A most recent analysis of the landscape of potential CT's  in \cite{FE} shows that many of them are  not ruled out on the basis of linearized Ward identities.

The infinite set of covariant CT's in \cite{Howe:1980th, Kallosh:1980fi} depends on constrained superfields describing the  superspace geometry.  These superfields satisfy the non-linear classical equations of motion. They cannot be used as unconstrained superfield variables in the path integral. The Lorentz covariant computations have to be done either using the gauge-fixed action of \cite{Cremmer:1979up} or using the unitarity cut method \cite{Dixon:2010gz}. In both of these cases, none of the 32 supersymmetries are manifest.

\N=8 supergravity has also a description in the LC real superspace with 16 Grassmann coordinates   based on  \cite{Brink:1982pd,Brink:2008qc}.
In this superspace  an unconstrained scalar chiral superfield and an anti-chiral one, related to the chiral one, are available. Therefore,  using the 1/2 of the manifest supersymmetry, one can  study the candidate CT's in the LC superspace where an unconstrained scalar chiral superfield,  describing the 256 propagating degree's of freedom,  serves as a variable in the path integral. The unpleasant  part of the LC superspace is the absence of the manifest Lorentz covariance. However, it is the only known superspace formalism that supports the path integral of the theory\footnote{The unconstrained harmonic superspace is not available for \N=8 supergravity.}
 which may lead to predictions being confirmed (in anomaly-free case) or  falsified (in case of anomalies) by perturbative computations.  It has been pointed out in \cite{Kallosh:2008mq} and recently confirmed  in \cite{Brink:2010cd} that {\it not a single viable  counterterm in the  real LC superspace  has  been constructed so far}. As the result, \N=8 supergravity may be viewed  as UV finite, until the real LC superspace CT's with 16 manifest kinematical supersymmetries are found, which are compatible with other known forms of the CT's.

There are two possible explanations for  the absence of the 3-loop UV divergence  \cite{Dixon:2010gz}  predicted by the
 linearized  $R^4$ counterterm of \N=8 \footnote{A complete non-linear  supersymmetric $R^4$ CT is known for \N=1 supergravity \cite{Deser:1977nt}. }supergravity \cite{ Kallosh:1980fi, Howe:1981xy}. 
One explanation was given in \cite{Kallosh:2009db}.  It says that the computation in the chiral LC superspace with 8 Grassmann coordinates, based on \cite{Brink:1982pd,Brink:2008qc,Mandelstam:1982cb,Belitsky:2004sc},  does not support this UV divergence since it is non-local in terms of chiral superfields. The second explanation is that the non-linear supersymmetric completion of the $R^4$ term requires the presence of the $\phi^2 R^4$ term, which breaks the continuous \E \,symmetry  \cite{Brodel:2009hu}. The analysis is based on string theory computation where both $R^4$ and $\phi^2 R^4$ term are present.  It is not surprising that both $R^4$ and $\phi^2 R^4$ terms are present in string amplitudes, demonstrating  explicitly that the continuous \E \,symmetry is broken.

 Meanwhile, if we take  the violation of the continuous \E \, symmetry
 as a reason to rule out the $R^4$ 3-loop UV divergence, it opens the door to further investigation of the consequences of the continuous \E \, symmetry. Note that   in the four-dimensional perturbative \N=8 supergravity  the conserved Noether current is available \cite{Cremmer:1979up, Brink:1979nt,Brink:2008qc,Kallosh:2008ic,Bianchi:2008pu,ArkaniHamed:2008gz,Kallosh:2008ru,Bossard:2010dq}.  This was the proposal made in  \cite{Kallosh:2009db,Bianchi:2009wj} where the issue of a  continuous \E \,symmetry of perturbative \N=8 broken down to a discrete \E($\mathbb{ Z}$) \,symmetry by non-perturbative effects in $d=4$ was discussed in detail.  The point was made there that the perturbative $d=4$ \N=8 supergravity may well be consistent with anomaly-free continuous \E \,symmetry. But the theory may  also be studied in the background of \N=8 BPS black holes, which are responsible for breaking continuous \E \,symmetry down to a discrete \E($\mathbb{ Z}$) \,symmetry in $d=4$.

 In addition to four-dimensional states of \N=8 supergravity, string theory states also include the 1/2 BPS states \cite{Green:2007zzb},
 which in $d=4$ are associated with 1/2 BPS extremal black holes. These singular black holes become massless only at the boundary of the moduli space. Therefore one can consistently exclude them from the spectrum of states of perturbative $d=4$ \N=8 supergravity and  expect that the continuous \E \,symmetry is preserved in perturbative theory   \cite{Bianchi:2009wj}.  It has been argued in \cite{Green:2007zzb} that 1/2 BPS states (Kaluza-Klein monopoles as (electrical) Kaluza--Klein modes of a dual
torus) do not decouple in string theory limit to $d=4$. The results of our  paper explain  why $R^4$ and  $\phi^2 R^4$ terms are present in string amplitudes, even in the $d=4$ limit: these terms originate from 1/2 BPS string states running in the loops of string theory and breaking \E ($\mathbb{ R}$) \,symmetry down to \E($\mathbb{ Z}$); they do not decouple in the $d=4$ limit. These $R^4$ and  $\phi^2 R^4$ terms are absent in $d=4$ \N=8 perturbative supergravity since there are no 1/2 BPS states in loops of perturbative \N=8 and UV divergences respect the  \E \, Noether current conservation \cite{Kallosh:2008ic}.
 For  the discrete \E($\mathbb{ Z}$) there is no Noether current. We will show below  that in absence of these 1/2 BPS states,  it is not possible to construct the  candidate CT's for the UV divergences, which are D-terms in real 16-dimensional LC superspace with 16 kinematical supersymmetries manifest, only the F-terms  are available. Therefore the UV finiteness prediction for $d=4$ \N=8 perturbative supergravity depends crucially on the fact that there is only one unconstrained  multiplet with 256 states (a chiral LC superfield, depending on 8 Grassmann variables), decoupled from the non-perturbative 1/2 BPS ones. This is the major difference between $d=4$ perturbative \N=8 supergravity and the 
$d=4$  limit of string theory.

Therefore, we will take an attitude that the continuous \E \,symmetry controls the quantum loops in the perturbative $d=4$ \N=8 supergravity. Even in this case, the puzzle of the UV properties remains: on the one hand,  the covariant CT's  in all higher loops respecting the continuous \E \,symmetry   are available in \cite{Howe:1980th, Kallosh:1980fi}.  On the other hand,  they may or may not exist in the LC superspace, as suggested in \cite{Kallosh:2008mq}.

 In the \N=4 Yang-Mills theory, as well as in \N=8 supergravity, the real superspace and the LC action were introduced by Brink, Lindgren and Nilsson   in \cite{Brink:1982pd,Brink:2008qc}. The chiral LC superspace and the action of the LC \N=4 Yang-Mills theory was introduced by Mandelstam in \cite{Mandelstam:1982cb}. The relation between these two was clarified for the Yang-Mills theory in \cite{Belitsky:2004sc} and more recently in \cite{Fu:2010qi}.  The study of the \N=8 chiral superspace path integral and its predictions was performed in  \cite{Kallosh:2009db}, where  the relation to on shell superamplitudes was also established. The study of the local in the chiral LC superspace  amplitudes in \cite{Kallosh:2009db} led us to conclude that there are no UV divergences below seven loop order, $L=7$. This effect has an interpretation \cite{Kallosh:2010mk} as a light-by-light type scattering effect in  supergraphs when an account is taken of the dynamical supersymmetry, which is non-manifest
in  LC superspace.   A conjecture was made in \cite{Kallosh:2009db} that if the additional information about the continuous \E \,symmetry is added to this chiral LC superspace analysis, some stronger restrictions on available CT's may be deduced, which  may  lead to the proof of the all-loop UV finiteness of the theory. A step in this direction was made in \cite{Fu:2010qi} when we have recently introduced  a new path integral for the \N=4 supersymmetric Yang-Mills theory. It originates from the real LC superspace action but leads to covariant answers for the superamplitudes via the chiral superspace action. 

It turned out that the most effective way to proceed with the analysis of  the candidate CT's,  is to use all three available superspaces: the Lorentz covariant on shell  superspace, the  real   LC superspace, and the chiral LC superspace. First, we will build sets of all possible counterterms in all three superspaces. Second, we will compare them. 
We will find out that these three sets of potential CT's have a vanishing intersection. In other words, none of the candidate CT's can belongs to all of these three sets simultaneously. This rules out all potential candidates.

\section{Non-linear CT's versus amplitudes}
In QFT there are two possibilities to make predictions about UV divergences, based on the path integral formalism.

In the background field method \cite{DeWitt:1967ub}, assuming  that the theory is free of anomalies, one can study the background invariant local CT's. This is equivalent to imposing the non-linear set of Ward identities. For example, in gravity one can look at the local CT's depending on curvature tensor and covariant derivatives of the form  $ \int d^4x D^n R^k(x)$ for UV divergences of 1PI diagrams, if one computes the path integral in background covariant gauges. This corresponds to a  simplest possible version of nonlinear Ward identities \cite{Kallosh:1974yh}.

 Alternatively, one can look at the amplitudes  which are non-local  since they involve 1PR diagrams. These amplitudes must satisfy the linearized form of the Ward Identities. For example, in the pure gravity case, each amplitude has to depend on the linearized version of derivatives of the curvature tensor. The total amplitude has a contribution from the local contact terms as well as from trees.
For example, the 4-point amplitude on shell, which does not depend on the choice of the gauge, has to be described by an effective action  in the Fourier space of the form
\be \int \prod_{i=1}^{i=4} d^4p_i  \, \delta^4(\sum p_i)  R^{\rm lin}(p_1)  R^{\rm lin}(p_2)  R^{\rm lin}(p_3)  R^{\rm lin}(p_4) A(p_1, p_2, p_3, p_4) \ .
\label{grav}\ee
Here  $A(p_1, p_2, p_3, p_4)$ is,  in general, a non-polynomial function of momenta, $R^{\rm lin}$ is a shortcut for the linearized curvature tensor $R_{\mu\nu\lambda\delta}^{\rm lin}$ satisfying the linearized equation $R_{\mu\nu}^{\rm lin}=0$.

In the case of pure gravity in the second loop approximation there is a unique candidate counterterm $R^3$ which does not vanish on shell, when the non-linear equations of motion $R_{\mu\nu}=0$ for the background field are  valid,  as shown in  \cite{Kallosh:1974yh,vanNieuwenhuizen:1976vb}. The existence of the corresponding UV divergence was inferred from explicit computations in \cite{Goroff:1985th}.
Each $R_{\mu\nu\lambda\delta}$ depends on linear, quadratic, etc., powers of the background field graviton $h_{\mu\nu}$. To get the amplitudes one has to use the expansion of the background field as an iterative solution of the field equation $R_{\mu\nu}(h)=0$.  In a symbolic form it  is $h(p)= h^{in}(p)+ {1\over p^2} V_3 (h^{in})^2+..., $ where $V_3$ is a 3-point vertex and $h^{in}_{\mu\nu}$ is a free graviton satisfying the free linear equation.
This expansion of the background field in terms of the free fields corresponds to replacing each of the background gravitons in $R^3$ by  an infinite tree \cite{DeWitt:1967ub,Kallosh:1974yh}. In particular there is a tree with 4 gravitons. Together with the local 4-graviton contact term in the expansion of $R^3$, this forms a 4-point gauge-independent on-shell amplitude satisfying Ward identities,  where each graviton in the linearized effective action (\ref{grav}) satisfies a free field equation.

In supersymmetric theories the term $R^3$ is ruled out \cite{Grisaru:1976nn}
and one starts the same analysis from the supersymmetric generalizations of the $R^4$ terms. Here  the more general local CT's of the form $D^n R^4$ by itself will produce the 4-point amplitudes,  which are the  linearized versions of the counterterm and correspond  to the case (\ref{grav}) where $A(p_1, p_2, p_3, p_4)$ is a strictly polynomial function of momenta. Thus learning the properties of the local 4-point amplitudes satisfying all relevant linearized Ward Identities is an important step in classification of possible candidate CT's. This process was initiated in
\cite{Kallosh:2008mq,Kallosh:2008ru,Kallosh:2009db} and significantly  developed in \cite{FE,Elvang:2010jv} based on the intense  studies of the superamplitudes in  \cite{Bianchi:2008pu,Elvang:2009wd}. In
\cite{Howe:2010nu} the clarification of some \N=8 CT's was recently presented
 from the point of view of representations of the superconformal group.

 If the 4-point CT's are absent, one may proceed with the analysis of the covariant 5-point amplitudes since the 5-point 1PR  terms from the 4-point UV divergences are absent. If there are no 5-point CT's, one proceeds with the covariant 6-point CT's, etc.


\section{From covariant to  chiral LC  CT's and back}
Using the chiral LC superfield path integral, we proved in \cite{Kallosh:2009db} that there are no UV divergences below $L=7$. Other approaches at present seem to be in agreement with this. The recent  superconformal approach (without using the harmonic superspace) in \cite{Howe:2010nu} and the  recent string theory  analysis in  \cite{Vanhove:2010nf}  suggest that L = 7 is the first UV dangerous point for $d=4$ \N=8 supergravity.

Here we first remind that the infinite sequence of $L\geq 8$ loop CT's in \N=8 supergravity, respecting all symmetries, including the continuous \E \,symmetry, was constructed in \cite{Howe:1980th, Kallosh:1980fi}.
The superspace invariants for  $L\geq 8$ are geometric and depend on the background supertorsion and supercurvature,
\be
\kappa^{2(L-1)} \int d^4 x \, d^{32} \theta\, {\rm Ber} \, E \ {\cal L} (T, R)\ ,
\label{covariant}\ee
 where ${\rm Ber} \,  E$ is the  Berezenian  (superspace vielbein)
 and ${\cal L} (T, R)$ is a superspace scalar constructed from supertorsion $T$ and supercurvature $R$.
At $L=7$ the corresponding background covariant CT is a full superspace volume:
\be
\kappa^{12} \int d^4 x \, d^{32} \theta \, {\rm Ber} \, E= \rm Vol
\label{volume}\ee
 At \N$=2$ the analogous volume of the full superspace vanishes \cite{Sokatchev:1980td}, so it is not obvious if (\ref{volume}) is a candidate CT in \N=8.
 Below we put together some useful information on linearized 7 loop CT's.

 \subsection{$L=7$ and \E}

In $L=7$ one can construct the linearized $n$-point   invariants \cite{Howe:1980th}, which have  32 supersymmetries unbroken, of the form
\be
\kappa^{12} \int d^4 x \, d^{32} \theta \, W^n  \ .
\label{n}\ee
Here $W^n$ represents an $SU(8)$-invariant product of $n$ self-dual scalar superfields satisfying linear equations of motion
\be
W_{abcd}(x, \theta)= \epsilon_{abcdefgh} \overline {W}^{efgh}(x, \theta)  \ .
\ee
The first component of the superfield $W_{abcd}(x, \theta)$ is a scalar field $W_{abcd}(x)$.
The curvature Weyl spinors $R_{\alpha\beta\gamma\delta}$ appear with 4 $\theta$'s and  $\bar R_{\dot \alpha\dot \beta\dot \gamma\dot \delta}$ appear with 4 $\bar \theta$'s, see \cite{Howe:1980th,Kallosh:1980fi} for details. Consider for example few cases of amplitudes which have all graviton contribution: 

a) The 4-point linearly supersymmetric invariant is using 16 $\theta$'s for each Weyl curvature and remaining 16 $\theta$'s give 8 more derivatives, sandwiched between curvature spinors. Symbolically,
\be
\kappa^{12} \int d^4 x \, d^{32} \theta \, W^4= \kappa^{12} \int d^4 x D^8 R^2 \bar R^2+...
\label{4point}\ee
where $...$ means the linear level supersymmetric partners of $D^8 R^2 \bar R^2$. The 4-point supersymmetric invariant (\ref{4point})  also has a continuous linear level \E \,symmetry. The linear variation of the superfield is
\be
\delta W_{abcd}(x, \theta)= \Sigma_{abcd}  \ ,
\ee
where $ \Sigma_{abcd}$ is a constant 70-dimensional parameter of \E \,. Under such a variation, the counterterm variation is
\be
\delta _{E_{7(7)}}^{ \rm lin} \int d^4 x \, d^{32} \theta \, W^4= 4 \int d^4 x \, d^{32} \theta \,  \Sigma_{abcd} (W^3)^{abcd}=0  \ .
\label{E774}\ee
It vanishes because this is a 3-point function where each field is on shell and massless. Thus, from the perspective of the Lorentz-covariant full linearized superspace invariants, nothing prevents the 7-loop UV divergence of the 4-point amplitude.

b) The 5-point linearly supersymmetric invariant uses 20 $\theta$'s for each Weyl curvature and remaining 12 $\theta$'s give 6 more derivatives, sandwiched between curvature spinors. But since the measure of integration has an equal amount of $\alpha$ and $\dot \alpha$ spinors, it is impossible to contract 5 Weyl curvatures with 6 derivatives, each having one $\alpha$ and one $\dot \alpha$ index. Thus the counterterm $D^6 R^5$ is ruled our by linearized supersymmetry. {\it Same for any other odd number of curvatures}.

c) The next case of interest is $D^4 R^6$,
\be
\kappa^{12} \int d^4 x \, d^{32} \theta \, W^6= \kappa^{12} \int d^4 x D^4 R^3 \bar R^3+...
\label{6point}\ee
Note that it comes out only for NMHV  since we have to contract the curvature spinors with equal number of $R$ and $\bar R$. One observes that under the linear  level \E \,symmetry it transforms as
\be
\delta _{E_{7(7)}}^{ \rm lin} \int d^4 x \, d^{32} \theta \, W^6= 6 \int d^4 x \, d^{32} \theta \,  \Sigma_{abcd} (W^5)^{abcd}\neq 0 \ .
\label{E776}\ee
This expression does not vanish, the linearized \E \,symmetry is broken. It rules out the independent supersymmetric counterterm $\kappa^{12} \int d^4 x \, d^{32} \theta \, W^6$.

d)  The next case of interest is $ R^8$
\be
\kappa^{12} \int d^4 x \, d^{32} \theta \, W^8= \kappa^{12} \int d^4 x  R^4 \bar R^4+...
\label{8point}\ee
Again, same number of left handed and right handed curvature spinors, or using the N$^k$MHV language, it is N$^2$MHV amplitude. As before, it is not invariant under  the linear  level \E \, symmetry since
\be
\delta _{E_{7(7)}}^{ \rm lin} \int d^4 x \, d^{32} \theta \, W^8= 8 \int d^4 x \, d^{32} \theta \,  \Sigma_{abcd} (W^7)^{abcd}\neq 0 \ .
\label{E778}\ee
Higher powers of curvature will not form 7-loop local CT's. However, there will be an infinite number of other linearized CT's which have up to 8 gravitons and many other fields in the amplitudes, as shown in (\ref{n}). All terms with $n> 4$ break the linear  level \E \,symmetry.

Analogous  results were obtained recently in \cite{FE,Elvang:2010jv} and  in \cite{Howe:2010nu} with regard to the gravitational part of the 7-loop CT's.

\subsection{4-point CT's}
The 4-point superamplitudes are always MHV. Therefore it is easy to start our program of comparing the Lorentz covariant and chiral LC superspace CT's with those predicted by the real LC superspace for the 4-point case.

 In \cite{Kallosh:2009db}, the 7-loop LC superfield 4-point counterterm invariant under 32 linearized supersymmetries was given by the following expression in the momentum superspace for the effective action:
\be
 W_{\rm 7-loop}^4  =  \prod_{i=1}^{4}  \left (\int d^4p_i \delta(p_i^2) d^8 \eta_i \,  \Phi (p_i, \eta_i)\right ) \, \delta^4 \left (\sum_m^4 p_m  \right)   \delta^{16} \left (\sum _k^4  \lambda^\alpha_i \eta_{ai} \right) {\cal P}_{\rm 7-loop} \ ,
\label{part}\ee
where
\be
{\cal P} _{\rm 7-loop}=
  \kappa^{12}\left( {[34 ]^4[ 12]^4 } + {[13 ]^4 [ 24]^4 }+ {[14 ]^4 [ 23]^4 }\right) \ .
\ee
Here $\Phi(p_i, \eta_i)$ are the on shell LC chiral superfields in a chiral superspace where they   depend  on 8 variables $\eta_{ia}$ at each of the 4 points. One can see, upon some Grassmann integration, that this LC superfield linearized counterterm is a partner of the covariant counterterm  $\int d^4 x \, d^{32} \theta \, W^4$ in (\ref{4point}). They have the same 4-graviton amplitude: $\int d^4 x D^8 R^2 \bar R^2$.
To see it from (\ref{part}) one can rewrite it in an equivalent form, using some identities, so that
\be
{\cal P} _{\rm 7-loop}\sim
  \kappa^{12}  (s^4 + t^4 + u^4)   {[34 ]^4\over \langle 12\rangle ^4 }  \ .
\ee
Now one may integrate over  8 $\eta_1$ and 8 $\eta_2$ using the delta function $\delta^{16} \left (\sum _k^4  \lambda^\alpha_i \eta_{ai}\right )$. This will bring a factor $\langle 12\rangle ^8$. The $\eta_3$ and $\eta_4$ integration is performed so that the corresponding $\eta$ factors come from the superfields. Altogether we find, upon integration, the following expression:
\be
 W_{\rm 7-loop}^4 \sim
  \prod_{i=1}^{4}  \left (\int d^4p_i \delta(p_i^2)   \right)  (s^4 + t^4 + u^4) \bar h(p_1) \bar h(p_2) h(p_3) h(p_4)  \langle 12\rangle ^4 [34 ]^4 +...
  \ee
This, in turn, is the same as
\be
 W_{\rm 7-loop}^4 \sim
  \prod_{i=1}^{4}  \left (\int d^4p_i \delta(p_i^2)   \right)  (s^4 + t^4 + u^4) R_{\alpha \beta \gamma\delta} (p_1) R^{\alpha \beta \gamma\delta} (p_2) \bar R_{\dot \alpha \dot \beta \dot \gamma \dot \delta} (p_3) \bar R^{\dot \alpha \dot \beta \dot \gamma \dot \delta} (p_4)+... ,
  \ee
as shown in \cite{Kallosh:2008mq,Kallosh:2009db}. In both cases,  the covariant CT (\ref{4point}), as well as  (\ref{part}),  have all 32 linearized supersymmetries. Together with the coinciding  gravitational part, this shows that in components, as functions of curvature, gravitino etc,  these are the same expressions.

Analogously, any other linearized covariant counterterm can be related to a LC superfield amplitudes. The important difference is that at the loop order less than 7, the LC superfield analogs of the covariant CT's are non-local, which rules them out. For example, for the 3-loop case we find a non-local amplitude
\be
{\cal P} _{{\rm 3-loop}}\sim
  \kappa^{4}  {[34 ]^4\over \langle 12\rangle ^4 }  \ .
\label{3loop}\ee
When performing the computation using the path integral in the LC chiral superspace, one expects to have  only local  UV divergences, which is not the case in  (\ref{3loop}), which rules out the 3-loop UV divergence \cite{Kallosh:2009db}.

But at and above $L=7$, the linearized supersymmetry alone, without a more intense use of the non-linear \E \,symmetry, does not help to rule out the corresponding UV divergences.

Now we switch back to the Grassmann coordinate space in (\ref{part}). It was explained in \cite{Kallosh:2009db} that for this purpose one should split the Lorentz covariant 16-component delta function into a product of two Lorentz non-covariant 8-component ones:
\be
 \delta^{16} \left (\sum _k^4  \lambda^\alpha_i \eta_{ai} \right)=  \delta^{8} \left (\sum _i^4  \lambda^1_i \eta_{ai} \right)  \delta^{8} \left (\sum _i^4  \lambda^2_i \eta_{ai} \right).
\label{structure}\ee
To use the  LC path integral \cite{Kallosh:2009db,Fu:2010qi}, we take $\lambda^1=\sqrt p_+$ and $\lambda^2= { p_{\bot}\over \sqrt p_+}$. The first 8-dimensional delta function, $\delta^{8} \left (\sum _i^4  \lambda^1_i \eta_{ai} \right)$,  where $\lambda^1  \eta_{a}=\sqrt p_+ \eta_a \Rightarrow {\partial \over \partial
\theta^a}$, allows to perform all $\eta$-integrations via a standard Fourier transform to coordinate superspace with 8 chiral Grassmann coordinates $\theta^a$. The second 8-dimensional delta function $\delta^{8} \left (\sum _i^4  \lambda^2_i \eta_{ai} \right)$, where $\lambda^2  \eta_{a}={ p_{\bot}\over \sqrt p_+}  \eta_a \Rightarrow  { p_{\bot}\over  p_+}{\partial \over \partial
\theta^a}$,  sprinkles various powers of the differential operator  ${ p_{\bot}\over  p_+}{\partial \over \partial
\theta^a}$ over each of the four superfields. As a result,  the total expression for (\ref{part}) becomes an integral over  $d^8 \theta$ of the superfields $\Phi(p_i, \theta)$ and $\left ({ p_{\bot i}\over p_{+i}} {\partial \over \partial \theta}\right)^m \Phi(p_i, \theta)$ times a polynomial function of momenta ${\cal P} _{\rm 7-loop}$. Here $m\leq 8$, and for each of the 8 $SU(8)$ indices, for example for $a=1$, there is just one power of $\left ({ p_{\bot i}\over p_{+i}} {\partial \over \partial \theta^{a=1}}\right)$,  which acts on all superfields  $\Phi(p_i, \theta) $, one  after another,  according to  $ \delta^8 \left(\sum _{i=1}^{i=4}  \lambda^2_i \eta_{ai} \right)$.

The explicit form  of the Grassmann Fourier transform  \cite{Kallosh:2009db}
from  $\eta_{ia}$ of each superfield to a common variable $\theta$ in coordinate space is rather involved, so we will not present is here. However, only its general properties described above will be required for the following analysis.

\subsubsection{$L\geq 8$, $n=4$}
The linearized covariant 4-point CT's in $L\geq 8$ are given by the  integrals over the  32-dimensional superspace of the $SU(8)$ invariants depending on 4 powers of the scalar superfield $W_{abcd}$ and its covariant superspace derivatives
\be
 \kappa^{2(L-1)} \int d^4 x \, d^{32 }\theta \, {\cal L} ( W, D_{\alpha \dot \alpha}  W, D_\alpha^a W, \bar  D_{\dot \alpha a} W, ...) \ .
\label{L4point}\ee
The dependence on scalar curvature always enters via the curvature Weyl tensors (two of them chiral and two anti-chiral) with some derivatives acting on them.

In case of superamplitudes and chiral LC superspace CT's
for all higher loops, as long as we look only at the 4-point local superamplitudes, we  always have an MHV case, the dependence on Grassmann variables is the same as in (\ref{part}):
\be
 W_{\rm L-loop}^4  =  \prod_{i=1}^{4}  \left (\int d^4p_i \delta(p_i^2) d^8 \eta_i \,  \Phi (p_i, \eta_i)\right ) \, \delta^4 \left (\sum_m^4 p_m  \right)   \delta^{16} \left (\sum _k^4  \lambda^\alpha_i \eta_{ai} \right) {\cal P}_{\rm L-loop}^4\ ,
\label{Lpart}\ee
where
\be
{\cal P} _{\rm L-loop}^4=
  \kappa^{2(L-1)}\left( {[34 ]^4[ 12]^4 } + {[13 ]^4 [ 24]^4 }+ {[14 ]^4 [ 23]^4 }\right) f(s,t,u)\ .
\label{LpartP}\ee
Here $f(s,t,u)$ is a polynomial function of the Mandelstam variables $s,t,u$ of dimension $2(L-3)$ symmetrised over all 4 momenta. The symmetrization actually comes in the effective action (\ref{Lpart}) automatically due to a symmetric dependence on scalar superfields in  (\ref{Lpart}).
We compare the results with the old Lorentz covariant CT's, in their linearized form.  We find  an agreement between (\ref{L4point}) and (\ref{Lpart}), (\ref{LpartP}).

\section{CT's in the real LC superspace}

Here we will explain the relation between the real and chiral
LC effective actions for \N=8 supergravity.  The classical action depends on {\it one unconstrained  scalar superfield}. It can be taken to be either  chiral $\phi$ or anti-chiral $\bar \phi$. In the real basis, the action  \cite{Brink:1982pd,Brink:2008qc} 
\be
S^{\rm real} [\phi, \bar \phi] ={1\over 2 \kappa^2}  \int d^4x \, d^8\theta \, d^8\bar \theta \,{\cal L}  (\phi, \bar \phi )
\label{action1}\ee
depends both on the chiral $\phi(x, \theta, \bar \theta)$ and anti-chiral $\bar \phi (x, \theta, \bar \theta)$ superfields and their supercovariant derivatives
\be
\bar d _a \phi \equiv\left ( {\partial \over \partial \theta^a} + {i \over \sqrt 2} \bar \theta_a \partial ^+ \right) \phi \, , \qquad
d ^a \bar \phi \equiv \left(- {\partial \over \partial \bar \theta_a} - {i \over \sqrt 2}  \theta^a \partial ^+ \right ) \phi \, , \qquad \{d^a, \bar d_b\}= -i\sqrt 2 \delta^a{}_b \partial^+ \bar \phi  \ .
\label{der}\ee
The action also has a polynomial dependence on the transverse space-time derivatives $ \bar \partial  $ and $ \partial $ and a non-polynomial dependence on $\partial^+  $.
The chirality condition is
\be
\bar d_a \bar \phi=0 \, , \qquad d^a \phi=0 \, .
\ee
 The anti-chiral field is not an independent one, as they both describe  a   single CPT invariant multiplet. The relation between them is given by
\be
\bar \phi= {1\over \partial ^{4}_+ } \bar d ^8  \phi \ .
\label{constr}\ee
The action (\ref{action1}) has 16 kinematical supersymmetries manifest in the real basis, where
\be
\delta \phi= (\bar \epsilon_a q^a  + \bar q_a \epsilon^a ) \phi \ , \qquad  \delta \bar \phi= (\bar \epsilon_a q^a  + \epsilon^a \bar q_a) \bar \phi \ .
\ee
Here the differential operators
\be
\bar q _a\equiv {\partial \over \partial \theta^a} - {i \over \sqrt 2} \bar \theta_a \partial ^+ \, , \qquad
q ^a \equiv - {\partial \over \partial \bar \theta_a} + {i \over \sqrt 2}  \theta^a \partial ^+ \, , \qquad \{q^a, \bar q_b\}= i\sqrt 2 \delta^a{}_b \partial^+
\ee
commute with covariant derivatives (\ref{der}). The action (\ref{action1}) has manifest kinematical supersymmetry associated with eight  $\theta$ and eight  $ \bar \theta$ coordinates of the real LC superspace. The other 16 dynamical supersymmetries, Lorentz symmetry and
\E \  symmetry, are realized non-linearly.

When $\bar \phi$ in the action is substituted by its expression in (\ref{constr}), one finds the chiral superspace action by integrating over the $\bar \theta$ variables:
\be
S^{\rm chiral} [\phi ] ={1\over 2 \kappa^2}  \int d^4x \, d^8\theta \left[ \int  d^8\bar \theta \,{\cal L} \,
(\phi, \bar \phi (\phi))\right] \ .
\label{action2}\ee
A generic feature of the chiral action above is: in the chiral basis  the chiral superfield is $\bar \theta$-independent since
\be
\bar d _a = {\partial \over \partial \theta^a} + { 2 i \over \sqrt 2} \bar \theta_a \partial ^+ \, , \qquad
d ^a = - {\partial \over \partial \bar \theta_a}  \, , \qquad \{d^a, \bar d_b\}= -i\sqrt 2 \delta^a{}_b \partial^+ \ ,
\label{der1}\ee
and
\be
d ^a \phi =  - {\partial \over \partial \bar \theta_a} \phi =0 \ .
\ee
The $d^8 \bar \theta$ integration  is straightforward, since the only source of  powers of $\bar \theta$ comes from $\bar d _a = {\partial \over \partial \theta^a} + { 2 i \over \sqrt 2} \bar \theta_a \partial ^+$ acting on chiral superfields. To get a $\theta$-derivative of the chiral superfield   we need to have an action ${\cal L}
(\phi, \bar \phi (\phi))$ depending on 4  superfields of the form
\be
\left (\bar d _{a=1} \phi_1 \right) \left(\bar d_{a=1} \phi_2 \right) \phi_3 \, \phi_4 = \left ( \left ( {\partial \over \partial \theta^{a=1}} + { 2 i \over \sqrt 2} \bar \theta_{a=1} \partial ^+ \right)   \phi_1\right)  \left ( \left ( {\partial \over \partial \theta^{a=1}} + { 2 i \over \sqrt 2} \bar \theta_{a=1} \partial ^+  \right) \phi_2 \right)  \phi_3 \, \phi_4 \ . ~~~
\label{simple}\ee
This will allow the comparison with the chiral LC superspace 4-point CT's in (\ref{part}) and  (\ref{Lpart}), see also the discussion around Eq.  (\ref{structure}).
Here the notation $\phi_i$ with different numbers  $i$ is used for  the chiral superfield $\phi$  hit by various other space-time derivatives. In (\ref{simple}) we keep track only of supercovariant derivatives in the $SU(8)$ direction $a=1$. The reason why we need two $\bar d _{a=1}$ derivatives in the action is the following. In (\ref{simple}) there are 2 terms linear in $\bar \theta_{a=1}$ which will give a non-vanishing chiral action upon $ \theta_{a=1}$ integration
\be
\int d\bar \theta_{a=1} \left [ \left (  \bar \theta_{a=1} \partial ^+   \phi_1\right)  \left (   \partial_{ \theta^{a=1}}   \phi_2 \right) + \left (  \partial _{\theta^{a=1}}   \phi_1 \right)  \left (  \bar \theta_{a=1} \partial ^+   \phi_2\right)\right  ] \phi_3 \, \phi_4  \ .
\label{simplest}\ee
The answer\footnote{The same structure was noticed  in Yang-Mills theory in \cite{Mandelstam:1982cb,Belitsky:2004sc}.}
 is proportional to
\be
\left[  \partial ^+   \phi_1 \; \partial_{ \theta^{a=1}}   \phi_2  -   \partial ^+
 \phi_2  \;  \partial_{  \theta^{a=1}}   \phi_1   \right] \phi_3 \, \phi_4  \ .
\label{Mandel}
\ee
If we would have a term in the action with one $\bar d _{a=1}$, we would get no $ {\partial \over \partial \theta^{a=1}}   \phi$ terms. If we would have 3 of them we would get two of  $ \partial_{  \theta^{a=1}}   \phi$
terms, but we need only one for (\ref{part}) and  (\ref{Lpart}), since the 4-point amplitudes are MHV, Grassmann degree 8. It will be different for N$^k$MHV CT's which have  higher  Grassmann degree for $k\geq 1$. We will return to this issue when considering CT's for $n$-point amplitudes with $n>4$. Finally, the term with 4 $\bar d _{a=1}$ is a total derivative. Therefore, the only possibility to have one spinorial derivative of the chiral superfield in a chiral action requires to have a term (\ref{simple}) in the original action and terms like (\ref{Mandel}) in the chiral action.

 The dependence on Grassmann variables in
the chiral LC superspace 4-point CT's in (\ref{part}) and  (\ref{Lpart}), as explained below  Eq.  (\ref{structure}), is
\be
\left ({\partial_{\bot} \over \partial^+}  \partial _{\theta^{a=1}}  \phi _1 \right ) \; \phi_2 \; \phi_3 \; \phi_4 +...
\label{exp}\ee
Here $\partial_{\bot}$ is a transverse space-time derivative.  Direct inspection shows that the functional dependence on Grassmann variables in the LC superfield amplitudes (\ref{Mandel}) originating from the real LC superspace action is incompatible with the one predicted for the CT's in chiral LC superspace of the form (\ref{exp}). It is even more convenient to compare these two functions of Grassmann variables in the momentum space following \cite{Fu:2010qi}, where the Fourier transform of (\ref{Mandel}) was described by the function
\be
\psi_{ij, a}\equiv \sqrt {p_{i +}}  \eta_{aj}- \sqrt {p_{j +}}   \eta_{ia} \ ,
\ee
to be compared with
\be
\sum _{i=1} ^{i=4}  \lambda^2_i \eta_{a i}  = \sum _{i=1} ^{i=4} \left ( { p_{\bot}\over \sqrt p_+}\right)_i \eta_{a i} \ .
\label{exp1}\ee
It was also shown in \cite{Fu:2010qi} that the dependence on $\psi_{ij, a}$ breaks the dynamical supersymmetry, whereas the CT's with (\ref{exp1}) inserted into (\ref{part}) and  (\ref{Lpart}) preserve it.

Thus we have shown here that the 4-point CT's from the chiral LC superspace, corresponding to   integrals over 8 $\theta$'s,   cannot be presented as real LC   local CT's, i.e. as integrals over 8 $\theta$'s and 8 $ \bar \theta$'s. They are true F-terms of the LC superspace and therefore do not lead to the UV divergences of the 4-point amplitudes at any loop order.

\section{$n$-point N$^k$MHV CT's, $n> 4$, $L\geq 7$, $k\geq 0$}
If at any loop order we would find no reason for  4-point CT's to vanish,  there would be no need to proceed to $n$-point CT's with  $n> 4$ since the UV divergence of the 4-point amplitudes would be bad enough.  But since no valid CT's are available at the level of a 4-point function in arbitrary loop order $L$, we have to proceed to higher point amplitudes.  In \cite{Kallosh:2009db} it was noticed that at the $n$-point level there is a delay of the linearized local CT's in the chiral superspace at least to the $L=n+3$ level. One can  combine this information with the \E \, symmetry to find the stronger restrictions for the  $n$-point LC chiral superspace CT's. However, here we proceed by comparing the CT's from the real and chiral LC superspaces.

For $n$-point,  $n> 4$, MHV $k=0$, $L\geq 7$ CT's, we have the following modification in our analysis. On the LC chiral superspace side we find
\be
 W_{\rm L-loop}^n  =  \prod_{i=1}^{n}  \left (\int d^4p_i \delta(p_i^2) d^8 \eta_i \,  \Phi (p_i, \eta_i)\right ) \, \delta^4 \left (\sum_m^n p_m  \right)   \delta^{16} \left (\sum _k^n  \lambda^\alpha_i \eta_{ai} \right) {\cal P}_{\rm L-loop}^n (p_i) \ ,
\label{Lpartn}\ee
where
$
{\cal P} _{\rm L-loop}^n$
is the function of the on-shell momenta $p_i$ which has a factor   $\kappa^{2(L-1)}$ in front. It respects the chirality structure of each  chiral superfield, which means that there are factors of $[ij]$. An explicit form of this function is not important for our purposes. The only important fact is that there is no new dependence on Grassmann variables $\eta_i$, it is all included into $\delta^{16} \left (\sum _k^n  \lambda^\alpha_i \eta_{ai} \right)$. Therefore on the chiral side the analysis is as for the 4-point case.

 In the real LC superspace the dependence on $\theta$-derivatives has to be
 \be
\left[  \partial ^+   \phi_1 \; \partial_{ \theta^{a=1}}   \phi_2  -   \partial ^+
 \phi_2  \;  \partial_{  \theta^{a=1}}   \phi_1   \right] \phi_3  \, \phi_4 ...\, \phi_n
\label{Mandeln}
\ee
to produce the linear dependence on $ \partial_{  \theta^{a=1}}   \phi$.
This implies, for any loop order, exactly the same incompatibility between chiral  LC superspace and  real LC superspace MHV CT's, as for the 4-point case.

For N$^k$MHV $n> 4$, $k\geq 1$, $L\geq 7$ 
the  CT's  are given by expression analogous to (\ref{Lpartn}) where however ${\cal P}_{\rm L-loop}^n (p_i)$ is replaced by ${\cal P} _{\rm L-loop}^n(p_i, \eta_i)$ which
has a  Grassmann degree $8k$ where $1 \leq  k  \leq n-4$. They are given by a function depending only on momenta $p_i$ times a function polynomial in $\eta_{ia}$. For our purposes the precise form of this function is required only with regard to its $\eta_{ia}$ dependence. Fortunately, it has been established in \cite{Elvang:2009wd} from the linear level supersymmetric Ward identities that all $n$-point $L$-loop N$^k$MHV superamplitudes depend on the  factors
\be
m_{ijk, a}\equiv [ij] \eta_{ka} + [jk]\eta_{ia} + [ki] \eta_{ja} \ .
\ee
They are supersymmetric with respect to all 32 supersymmetries. Translating this to the LC variables in notation of
\cite{Fu:2010qi},
we have
\be
[p_i, p_j]=[p q]= { p_+ \bar q_{\bot} - q_+ \bar p_{\bot}\over \sqrt {p_+} \sqrt {q_+}} \ .
\ee
There are also more general structures in the real LC candidate CT's, since now for any direction in the $SU(8)$ space, like $a=1$, there will be higher powers of $ \partial_{  \theta^{a=1}}   \phi$.  For each $k$ there is an extra power of $ \partial_{  \theta^{a=1}}   \phi$. This means that we have to start with the more general expression  of the form
\be
\left (\bar d _{a=1} \phi_1 \right) \left(\bar d_{a=1} \phi_2 \right)... \left(\bar d_{a=1} \phi_{k+2} \right)  \phi_{k+3}...\phi_n   \ .
\label{notsimple}\ee
Integrating out all terms linear in $\bar \theta_{a=1}$, as before,  we find
\be
\epsilon ^{i_1 i_2 ...i_{k+2} } (\partial^+ \phi_{i_1} ) ( \partial_{\theta_{a=1}}\phi_{i_2} ) ... (\partial_{\theta_{a=1}}\phi_{i_{k+2}})\phi_{k+3}...\phi_n    = \epsilon ^{i_1 i_2 ...i_{k+2} } \psi_{i_1 i_2 ...i_{k+2}} \phi_{k+3} ...\phi_n \ .
\label{psi}\ee
In MHV $k=0$ case, this is a structure, described previously as $\epsilon ^{i j } (\partial^+ \phi_{i} ) ( \partial_{\theta_{a=1}}\phi_{j} ) \phi_{3}...\phi_n    =  \psi_{12} \phi_{3} ...\phi_n $.
This is to be compared with (\ref{Lpartn}), where
\be
{\cal P} _{\rm L-loop}^n \sim  \sum _{i=1} ^{i=n} \left ( { p_{\bot}\over \sqrt p_+}\right)_i \eta_{a=1 i}  \cdot X \ ,
\label{exp-n}
\ee
where $X$ is a product of $m_{ijk, a=1}$.  One can see that the CT's in (\ref{notsimple}), (\ref{psi}) and in (\ref{Lpartn}), (\ref{exp-n})
are incompatible, either by a direct inspection of by computing the variation of the real LC superspace CT's under dynamical supersymmetry: they maintain the kinematical symmetry off shell, but they are breaking linear level dynamical symmetry
$
\delta \phi = \epsilon^a \left ( { p_{\bot}\over \sqrt p_+}\right) \bar q_a \phi
$.
This finalizes the proof that all linearized on shell CT's with 32 unbroken supersymmetries are forbidden, because  they cannot appear in the  perturbative path integral in the real  LC superspace.


\section{Summary}
 30 years ago,  \N=8 supergravity in four dimensions was suspected  to be UV divergent at higher loop orders. The suspicion  was  based on a construction of an infinite set of superinvariants \cite{Howe:1980th,Kallosh:1980fi} in a Lorentz covariant on shell superspace geometry with 32 Grassmann coordinates. These counterterms  were viewed as candidates for UV divergences. 2 years ago,  one of the accusers  proposed to restore the presumption of  innocence, and proposed  \cite{Kallosh:2008mq} to check the prediction of the on shell full superspace \cite{Howe:1980th,Kallosh:1980fi} using the unconstrained off shell real and chiral LC superspaces.

The reason for this was the following. It was known that the full superspace has only on shell superfields, which cannot serve as variables in the path integral. 
It has been realized,  thanks to the progress with 3- and 4-loop computations \cite{Dixon:2010gz} and the general understanding of the  superamplitude structures of  \N=8 supergravity \cite{Bianchi:2008pu}, that things do not quite go in the direction expected in \cite{Kallosh:1980fi}. Trying to understand this discrepancy, we came to the conclusion that one should use the LC formalism \cite{Brink:1982pd,Brink:2008qc} of \N=8 supergravity. LC superspace has two versions: one  is the real superspace \cite{Brink:1982pd,Brink:2008qc} with 16 Grassmann coordinates.  The other one is the  chiral one \cite{Mandelstam:1982cb,Kallosh:2009db} with 8 Grassmann coordinates.
Since in both cases in \N=8 supergravity,  there is an {\it unconstrained chiral superfield with 256 propagating degrees of freedom},  one can construct the path integral, and analyze the consequences corresponding to Ward identities. In the chiral  LC superspace,  this was done in \cite{Kallosh:2009db,Fu:2010qi}. An infinite set of candidate CT's, respecting  linearized 32 supersymmetries and corresponding to those in the Lorentz covariant full superspace  \cite{Howe:1980th,Kallosh:1980fi}, was discovered in \cite{Kallosh:2009db}. The local CT's in the chiral LC analysis were shown to be absent for $L<7$.

Recent progress  in understanding superamplitudes led to a systematic analysis of CT's of \N=8 supergravity in   \cite{FE}.  Using the relation between superapmlitudes in \cite{Bianchi:2008pu} and light-cone superfield amplitudes in  \cite{Kallosh:2009db},  one also finds systematic  information on the chiral LC superspace  candidate CT's.

With all this information at hand, time was ripe to look at the real LC superspace CT's. The relevant CT's must preserve {\it off shell  the  16 kinematic supersymmetries manifest  in the process of the supergraph computations in the real superspace}. This means that the rules are simple. The local CT's may depend on a chiral superfield and related  anti-chiral superfield, and their supercovariant derivatives, as well as positive and negative powers of  $p_+$ and positive powers of $ p_{\bot},  \bar p_{\bot}$. This dictates the form of the CT's in the Grassmann 8-dimensional chiral superspace, which originate from the real Grassmann 16-dimensional  LC superspace, when the integration over 8 $\bar \theta$ coordinates is performed. First, one has to replace every anti-chiral superfield, using the CPT conjugation,  by its expression  in terms of the chiral one.
Now in the chiral basis for the superfields the only dependence on $\bar \theta$  enters via the supercovariant derivatives. The integration $d^8 \bar \theta$ is straightforward; it picks up 8 powers of $\bar \theta$'s from supercovariant derivatives. This leads to a possibility to have factors ${\partial\over \partial \theta} $ acting on  chiral superfields $\phi(x, \theta)$ in the chiral LC superspace, but only in a particular combination. In the LC Yang-Mills theory in \cite{Mandelstam:1982cb} the relevant anti-commuting derivative was given a name: $\partial_{x^-, \theta}$.
For \N=8 supergravity,  the analogous anti-commuting derivative is available, in notation of  \cite{Fu:2010qi} it is simple in a Fourier superspace:  $\psi_{ij,a}=  \sqrt {p_{i +}}  \eta_{ja}- \sqrt {p_{j +}}   \eta_{ia}$. Thus, if we were able to construct the CT's in the real LC superspace with 16 Grassmann coordinates   we would be able to integrate them out over the 8 powers of $\bar \theta$. This would result in the appearance of the Grassmann variables in the chiral linearized LC counterterms of the form
$\psi_{ij,a}$. However, we now know that the chiral LC superspace linearized CT's may depend only on the following combination of Grassmann parameters: For the MHV case it is $ \delta^{8} \Bigl(\sum _i  {p_{\bot}\over \sqrt{p^+}} \eta_{ai} \Bigr)$  \cite{Kallosh:2009db}, for the N$^k$MHV case we have in addition the supersymmetric  factors of $m_{ijk,a}= ([ij]\eta_k+ [jk]\eta_i+[ki]\eta_j )_a$, as shown in   \cite{Elvang:2009wd}. These functions are incompatible with $\psi_{ij}, \psi_{ikj...}$, which are permitted when integrating out the $\bar \theta$-variables. One can  check that the dependence on $\psi_{ij}$ in contact terms in the LC action breaks the dynamical supersymmetry  \cite{Fu:2010qi},
 whereas the LC superspace covariant CT's have all 32 linearized supersymmetries intact.

After the use of these rather technical  tools with LC and Grassmann variables in CT's, we may summarize the situation in simple terms. If one choses to use the supergraph rules of the  chiral LC superspace with 8 Grassmann coordinates, the single 256 dimensional  multiplet of \N=8 supergravity  running in the loops of $d=4$ LC  supergraphs
leads to an infinite set of  candidate  CT's, compatible with the covariant ones. However,  each of  these CT's is an F-term of the full real LC superspace, which has 16 Grassmann coordinates.  The technical paragraph above shows that these F-terms cannot be presented as local D-terms in a real LC superspace with 16 manifest kinematic supersymmetries. This means that using the supergraph rules of the type  following from the action
 in  \cite{Brink:1982pd,Brink:2008qc},  one cannot build the  CT's compatible with the ones already known from the chiral  LC superspace and from the covariant  linearized Grassmann 32-dimensional superspace. Therefore all CT's are ruled out. The basis for this conclusion is the equivalence theorem for \N=8 supergravity \cite{Kallosh:2009db}. It is valid only in the anomaly-free case when the formal predictions from the path integral are confirmed by loop computations. From this perspective, the  future higher loop computations developing on \cite{Dixon:2010gz} will test the absence of quantum anomalies in the path integral and, hopefully, confirm the prediction made in \cite{Bossard:2010dq} about the absence of \E \, anomalies to all orders in perturbation theory.

Thus,  in this paper,  \N=8\, $d=4$ supergravity is  acquitted from the previous accusation in \cite{Howe:1980th,Kallosh:1980fi} and is predicted  to be UV finite if there are no anomalies violating  the equivalence theorem for physical observables. The observables are expected to be the same whether 
computed  using a Lorentz covariant  path integral for component fields based on a properly gauge-fixed Cremmer-Julia action  with the conserved Noether current of the continuous \E \,symmetry,  using the unitarity cut method, or using the  chiral or real LC supergraph Feynman rules.


\section*{Acknowledgments}
We are grateful to  M. Bianchi, L. Brink, S. Ferrara, D. Freedman, C. H. Fu, H. Elvang, P. Howe, G. Korchemsky, A. Linde, B. Nilsson, P. Ramond, K. Stelle, J. Schwarz, S. Shenker, E. Sokatchev,  and L. Susskind  for discussions.
This work  is
supported by the NSF grant 0756174.



\begin{thebibliography}{10}


\bibitem{Cremmer:1979up}
  E.~Cremmer and B.~Julia,
``The SO(8) Supergravity,''
  Nucl.\ Phys.\  B {\bf 159}, 141 (1979);
  B.~de Wit and H.~Nicolai,
 ``N=8 Supergravity,''
  Nucl.\ Phys.\  B {\bf 208}, 323 (1982);
  B.~de Wit and D.~Z.~Freedman,
 ``On SO(8) Extended Supergravity,''
  Nucl.\ Phys.\  B {\bf 130}, 105 (1977).


\bibitem{Brink:1979nt}
  L.~Brink and P.~S.~Howe,
 ``The \N=8 Supergravity In Superspace,''
  Phys.\ Lett.\  B {\bf 88}, 268 (1979).

\bibitem{Dixon:2010gz}
  L.~J.~Dixon,
``Ultraviolet Behavior of N=8 Supergravity,''
  arXiv:1005.2703 [hep-th];
  Z.~Bern, J.~J.~M.~Carrasco and H.~Johansson,
  ``Progress on Ultraviolet Finiteness of Supergravity,''
  arXiv:0902.3765 [hep-th].

\bibitem{Bossard:2010dq}
  G.~Bossard, C.~Hillmann and H.~Nicolai,
``Perturbative quantum E7(7) symmetry in N=8 supergravity,''
  arXiv:1007.5472 [hep-th];
  N.~Marcus,
``Composite Anomalies In Supergravity,''
  Phys.\ Lett.\  B {\bf 157}, 383 (1985);
  P.~di Vecchia, S.~Ferrara and L.~Girardello,
``Anomalies Of Hidden Local Chiral Symmetries In Sigma Models And Extended
 Supergravities,''
  Phys.\ Lett.\  B {\bf 151}, 199 (1985);
  L.~Alvarez-Gaume and E.~Witten,
  ``Gravitational Anomalies,''
  Nucl.\ Phys.\  B {\bf 234}, 269 (1984);
  O.~Alvarez, I.~M.~Singer and B.~Zumino,
``Gravitational Anomalies And The Family's Index Theorem,''
  Commun.\ Math.\ Phys.\  {\bf 96}, 409 (1984).





\bibitem{Kallosh:2009db}
  R.~Kallosh,
  ``N=8 Supergravity on the Light Cone,''
  Phys.\ Rev.\  D {\bf 80}, 105022 (2009)
  [arXiv:0903.4630 [hep-th]].


\bibitem{Howe:1980th}
  P.~S.~Howe and U.~Lindstrom,
  ``Higher Order Invariants In Extended Supergravity,''
  Nucl.\ Phys.\  B {\bf 181}, 487 (1981).


\bibitem{Kallosh:1980fi}
  R.~E.~Kallosh,
``Counterterms in extended supergravities,''
  Phys.\ Lett.\  B {\bf 99} (1981) 122.

\bibitem{Kallosh:2008mq}
  R.~Kallosh,
``On a possibility of a UV finite N=8 supergravity,''
  arXiv:0808.2310 [hep-th].



\bibitem{Bossard:2009sy}
  G.~Bossard, P.~S.~Howe and K.~S.~Stelle,
  ``The ultra-violet question in maximally supersymmetric field theories,''
  Gen.\ Rel.\ Grav.\  {\bf 41}, 919 (2009)
  [arXiv:0901.4661 [hep-th]].

    \bibitem{FE}  Talks by D. Freedman and H. Elvang  at the Ahoop conference: \\
 http://people.physik.hu-berlin.de/$\sim$ahoop/program.shtml




\bibitem{Brink:1982pd}
  L.~Brink, O.~Lindgren and B.~E.~W.~Nilsson,
  ``N=4 Yang-Mills Theory On The Light Cone,''
  Nucl.\ Phys.\  B {\bf 212}, 401 (1983);
  A.~K.~H.~Bengtsson, I.~Bengtsson and L.~Brink,
``Cubic Interaction Terms For Arbitrarily Extended Supermultiplets,''
  Nucl.\ Phys.\  B {\bf 227}, 41 (1983);
  S.~Ananth, L.~Brink and P.~Ramond,
  ``Eleven-dimensional supergravity in light cone superspace,''
  JHEP {\bf 0505}, 003 (2005)
  [arXiv:hep-th/0501079].
  S.~Ananth, L.~Brink, R.~Heise and H.~G.~Svendsen,
  ``The N=8 Supergravity Hamiltonian as a Quadratic Form,''
  Nucl.\ Phys.\  B {\bf 753}, 195 (2006)
  [arXiv:hep-th/0607019].



\bibitem{Brink:2008qc}
  L.~Brink, S.~S.~Kim and P.~Ramond,
``$E_{7(7)}$ on the Light Cone,''
  JHEP {\bf 0806}, 034 (2008)
  [arXiv:0801.2993 [hep-th]].


\bibitem{Brink:2010cd}
  L.~Brink,
  ``Maximal supersymmetry and exceptional groups,''
  arXiv:1006.1558 [hep-th].



\bibitem{Howe:1981xy}
  P.~S.~Howe, K.~S.~Stelle and P.~K.~Townsend,
  ``Superactions,''
  Nucl.\ Phys.\  B {\bf 191}, 445 (1981).


\bibitem{Deser:1977nt}
  S.~Deser, J.~H.~Kay and K.~S.~Stelle,
``Renormalizability Properties Of Supergravity,''
  Phys.\ Rev.\ Lett.\  {\bf 38}, 527 (1977).



\bibitem{Mandelstam:1982cb}
  S.~Mandelstam,
 ``Light Cone Superspace And The Ultraviolet Finiteness Of The N=4 Model,''
  Nucl.\ Phys.\  B {\bf 213}, 149 (1983);
  L.~Brink, O.~Lindgren and B.~E.~W.~Nilsson,
  ``The Ultraviolet Finiteness Of The N=4 Yang-Mills Theory,''
  Phys.\ Lett.\  B {\bf 123}, 323 (1983).



\bibitem{Belitsky:2004sc}
  A.~V.~Belitsky, S.~E.~Derkachov, G.~P.~Korchemsky and A.~N.~Manashov,
``Dilatation operator in (super-)Yang-Mills theories on the light cone,''
  Nucl.\ Phys.\  B {\bf 708} (2005) 115
  [arXiv:hep-th/0409120].









\bibitem{Brodel:2009hu}
  J.~Broedel and L.~J.~Dixon,
 ``$R^4$ counterterm and E7(7) symmetry in maximal supergravity,''
  JHEP {\bf 1005}, 003 (2010)
  [arXiv:0911.5704 [hep-th]];
  H.~Elvang and M.~Kiermaier,
  ``Stringy KLT relations, global symmetries, and $E_7(7)$ violation,''
  arXiv:1007.4813 [hep-th].

\bibitem{Bianchi:2008pu}
  M.~Bianchi, H.~Elvang and D.~Z.~Freedman,
  ``Generating Tree Amplitudes in N=4 SYM and N = 8 SG,''
  JHEP {\bf 0809}, 063 (2008)
  [arXiv:0805.0757 [hep-th]].


\bibitem{Kallosh:2008ic}
  R.~Kallosh and M.~Soroush,
  ``Explicit Action of E7(7) on N=8 Supergravity Fields,''
  Nucl.\ Phys.\  B {\bf 801}, 25 (2008)
  [arXiv:0802.4106 [hep-th]].
  ``The footprint of E7 in amplitudes of N=8 supergravity,''
  JHEP {\bf 0901}, 072 (2009)
  [arXiv:0811.3414 [hep-th]].


\bibitem{ArkaniHamed:2008gz}
  N.~Arkani-Hamed, F.~Cachazo and J.~Kaplan,
 ``What is the Simplest Quantum Field Theory?,''
  arXiv:0808.1446 [hep-th].

\bibitem{Kallosh:2008ru}
  R.~Kallosh, C.~H.~Lee and T.~Rube,
  ``N=8 Supergravity 4-point Amplitudes,''
  JHEP {\bf 0902}, 050 (2009)
  [arXiv:0811.3417 [hep-th]].



\bibitem{Bianchi:2009wj}
  M.~Bianchi, S.~Ferrara and R.~Kallosh,
  ``Perturbative and Non-perturbative N =8 Supergravity,''
  Phys.\ Lett.\  B {\bf 690}, 328 (2010)
  [arXiv:0910.3674 [hep-th]];
  M.~Bianchi, S.~Ferrara and R.~Kallosh,
 ``Observations on Arithmetic Invariants and U-Duality Orbits in N =8
Supergravity,''
  JHEP {\bf 1003}, 081 (2010)
  [arXiv:0912.0057 [hep-th]].


\bibitem{Green:2007zzb}
  M.~B.~Green, H.~Ooguri and J.~H.~Schwarz,
  Phys.\ Rev.\ Lett.\  {\bf 99}, 041601 (2007)
  [arXiv:0704.0777 [hep-th]].



\bibitem{Kallosh:2010mk}
  R.~Kallosh and P.~Ramond,
 ``Light-by-Light Scattering Effect in Light-Cone Supergraphs,''
  arXiv:1006.4684 [hep-th].

\bibitem{Fu:2010qi}
  C.~H.~Fu and R.~Kallosh,
 ``New N=4 SYM Path Integral,''
  arXiv:1005.4171 [hep-th].



\bibitem{DeWitt:1967ub}
  B.~S.~DeWitt,
 ``Quantum theory of gravity. II. The manifestly covariant theory,''
  Phys.\ Rev.\  {\bf 162}, 1195 (1967).

\bibitem{Kallosh:1974yh}
  R.~E.~Kallosh,
  ``The Renormalization In Nonabelian Gauge Theories,''
  Nucl.\ Phys.\  B {\bf 78}, 293 (1974).



\bibitem{vanNieuwenhuizen:1976vb}
  P.~van Nieuwenhuizen and C.~C.~Wu,
  ``On Integral Relations For Invariants Constructed From Three Riemann Tensors
  And Their Applications In Quantum Gravity,''
  J.\ Math.\ Phys.\  {\bf 18}, 182 (1977).

\bibitem{Goroff:1985th}
  M.~H.~Goroff and A.~Sagnotti,
  Nucl.\ Phys.\  B {\bf 266}, 709 (1986).
  A.~E.~M.~van de Ven,
  ``Two loop quantum gravity,''
  Nucl.\ Phys.\  B {\bf 378}, 309 (1992).

\bibitem{Grisaru:1976nn}
  M.~T.~Grisaru,
 ``Two Loop Renormalizability Of Supergravity,''
  Phys.\ Lett.\  B {\bf 66}, 75 (1977).


\bibitem{Elvang:2010jv}
  H.~Elvang, D.~Z.~Freedman and M.~Kiermaier,
 ``A simple approach to counterterms in N=8 supergravity,''
  arXiv:1003.5018 [hep-th].


\bibitem{Elvang:2009wd}
  H.~Elvang, D.~Z.~Freedman and M.~Kiermaier,
  ``Solution to the Ward Identities for Superamplitudes,''
  arXiv:0911.3169 [hep-th].



\bibitem{Howe:2010nu}
  J.~M~Drummond, P.~J.~Heslop, P.~S.~Howe
  ``A note in N=8 counterterms,''
  [arXiv:1008.4939 [hep-th]].

  \bibitem{Vanhove:2010nf}
  P.~Vanhove,
 ``The critical ultraviolet behaviour of N=8 supergravity amplitudes,''
  arXiv:1004.1392 [hep-th].
  J.~Bjornsson and M.~B.~Green,
``5 loops in 24/5 dimensions,''
  arXiv:1004.2692 [hep-th].

\bibitem{Sokatchev:1980td}
  E.~Sokatchev,
 ``A Superspace Action For N=2 Supergravity,''
  Phys.\ Lett.\  B {\bf 100}, 466 (1981).





\end{thebibliography}
\end{document}